\documentclass [a4paper]{article}
\title{Simple applications of Noether's first theorem in quantum mechanics and electromagnetism}
\author{Harvey R Brown\thanks{Faculty of Philosophy, University of Oxford, 10 Merton Street, Oxford OX2 7TL, U.K.; {\em harvey.brown@philosophy.ox.ac.uk.}}   \and Peter Holland\thanks{Green College, University of Oxford, Woodstock Road, Oxford OX2 6HG; {\em peter.holland@green.ox.ac.uk.}}}
\begin{document}
\maketitle
\begin{abstract}
Internal global symmetries exist for the free non-relativistic Schr\"{o}dinger particle, whose associated Noether charges---the space integrals of the wavefunction and the wavefunction multiplied by the spatial coordinate---are exhibited. Analogous symmetries in classical electromagnetism are also demonstrated.
\end{abstract}
\section{Introduction}

A deep insight in physics concerns the connection between continuous symmetry principles of the `global' variety and conservation laws, famously treated---though not discovered---by Emmy Noether in her 1918 work\footnote{See Noether~\cite{noether}; an English translation is found in Tavel~\cite{trans}, and a nice historical analysis of Noether's 1918 work in Kastrup~\cite{kas:his}.}  in the calculus of variations. Noether's study of course dealt with dynamical systems susceptible to a Lagrangian formulation, and in 1954, Eugene Wigner warned against a ``facile identification'' of symmetry and conservation principles, his main argument being that not all interesting dynamical systems are of this kind.\footnote{See Wigner~\cite{wigner}.}

It is worth emphasizing that even when the dynamics \textit{are} Lagrangian--which is ususally the case---the significance of Noether's `first' theorem, the one dealing with global symmetries, can be quite a subtle issue. First, not all such symmetries are Noetherian, in the sense of meeting the conditions normally laid down for a Noether-type theorem to hold (examples are found in section 4 below). Second, even when the symmetries are Noetherian, there are ways in which they may not lead to equations of continuity at all---let alone conserved charges.\footnote{See, for example, Trautman~\cite{traut} sections 5-2 and 5-3, and Brown and Brading~\cite{bb} section II.} But, most significantly for our purposes, even when a connection with continuity equations does hold, the following points are pertinent.

(i) The continuity equations may be trivial in the light of the equations of motion (the Euler-Lagrange equations): the two sets of equations may indeed coincide.

(ii) The continuity equation may not lead to the existence of conserved Noether charges.

(iii) Neither the Lagrangian nor the conserved charges need be real-valued.

(iv) The Noether symmetry transformation may not have an 'active' interpretation. In particular it may not carries states of the dynamical system into states.

The main point of this paper is to illustrate points (i) to (iv) by way of simple examples involving the non-relativistic Schr\"{o}dinger dynamics of a free particle and classical electrodynamics. It is noteworthy that the full space-time symmetry group associated with the former was only demonstrated many years after the development of quantum mechanics.\footnote{See Niederer~\cite{nied} and Jahn and Sreedhar~\cite{js}.} In the present note, the Noether charges associated with unfamiliar \emph{internal} symmetries will be exhibited---one of these symmetries being a non-relativistic analogue of a non-unitarily implementable Noetherian symmetry in relativistic field theory. 

\section{Noether's first theorem for internal symmetries}

We begin by considering a Lagrangian density $\mathcal{L}$ depending on a system of $N$ non-relativistic fields $\psi_{k}$ ($k=1, \ldots ,N$) and their first derivatives. The Lagrangian $L$ for this system of fields, defined relative to some (possibly time-dependent) compact, bounded region of space $\Omega$, is
\begin{equation}
L = \int_{\Omega} \mathcal{L}d\mathbf{x}.    \label{eq:lag}
\end{equation}
and the action defined relative to $\Omega$ and the two times $t_{1}$ and $t_{2}$ is
\begin{equation}
S = \int_{t_{1}}^{t_{2}}\!\!\int_{\Omega}\mathcal{L}d\mathbf{x}dt.    \label{eq:action}
\end{equation}
We refer to the space-time region defined by $R$, $t_{1}$ and $t_{2}$ as $\Sigma$ and simply write
\begin{equation}
S = \int_{\Sigma}\mathcal{L}d\mathbf{x}dt.    \label{eq:simple-action}
\end{equation}
If now we introduce infinitesimal variations in \textit{both} the fields $\psi_{k}$ (the dependent variables) and the coordinates $\mathbf{x}, t$ (the independent variables), then the corresponding first-order variation in $S$ can be written as the sum of two terms. The first, still an integral over $\Sigma$, has as its integrand the sum 
$\sum_{k} \mathcal{E}_{\psi_{k}}\bar{\delta}\psi_{k}$ where $\bar{\delta}\psi_{k}$ denotes the `total' variation, or Lie drag of $\psi_k$ (the definition of which need not concern us), where the $\mathcal{E}_{\psi_{k}}$ are the Euler expressions
\begin{equation}
\mathcal{E}_{\psi_{k}} = \frac{\partial \mathcal{L}}{\partial\psi_{k}} - \mathbf{\nabla}\cdot\frac{\partial \mathcal{L}}{\partial (\mathbf{\nabla} \psi_{k})} - \partial_{t}\frac{\partial \mathcal{L}}{\partial \dot{\psi_{k}}},  \label{euler-expression}
\end{equation}
and where $\partial_t = \partial/\partial t$ and $\dot{\psi_k} = \partial\psi_k/\partial t$. The second term in the first variation of $S$ takes the form of the integral of a divergence. Hence by Gauss's theorem it is a surface term, i.e. an integral over the three-dimensional boundary $\Gamma (\Sigma)$ of the space-time region $\Sigma$.

If we apply Hamilton's principle in relation to arbitrary infinitesimal variations of a \textit{specific} field $\psi_{k}$ that vanish on the boundary $\Gamma (\Sigma)$, this means that the action is stationary under such variations for arbitrary choice of the spacetime region $\Sigma$.\footnote{For a detailed treatment of Hamilton's principle in the context of non-relativistic fields, see Doughty~\cite{doughty}, pp. 207--209, Cohen-Tannoudji \textit{et al}.\ ~\cite{cdg} pp. 92-93, and Jos\'{e} and Saletan~\cite{js*} pp. 557--561.} So the second term in the first variation of $S$ automatically vanishes and as is well-known, for the first term to vanish on its own it must be the case that the Euler-Lagrange equation for $\psi_{k}$ holds:
\begin{equation}
\mathcal{E}_{\psi_{k}} = \frac{\partial \mathcal{L}}{\partial\psi_{k}} - \mathbf{\nabla}\cdot\frac{\partial \mathcal{L}}{\partial (\mathbf{\nabla} \psi_{k})} - \partial_{t}\frac{\partial \mathcal{L}}{\partial \dot{\psi_{k}}} = 0.  \label{eq:euler-lagrange}
\end{equation}

Noether's concern in 1918 was with a different stationarity problem, one defined for \textit{specific} infinitesimal transformations in (possibly) both the dependent and the independent variables, which do not necessarily vanish on the boundary $\Gamma (\Sigma)$ (so in the first instance both of the above-mentioned terms feature in the first variation of the action). Noether was interested in those transformations for which the first variation in the action strictly vanishes, but since 1918 it has long been recognised that ``quasi-invariance'' of the action is sufficient for the transformations in question to count as \textit{symmetry} transformations. By this is meant that the corresponding first variation in the Lagrangian density vanishes only up to a `four-divergence' term:
\begin{equation}
\delta \mathcal{L} = \mathbf{\nabla}\!\cdot\!\mathbf{\Lambda} + \frac{d\Lambda^{0}}{dt},   \label{eq:quasi-invariance}
\end{equation}
where $\mathbf{\Lambda}$ and $\Lambda^{0}$ are some infinitesimal vector and scalar fields respectively,\footnote{A common misunderstanding about the nature of $\mathbf{\Lambda}$ and $\Lambda^{0}$ is discussed in point (ii) in the Appendix below.} and the derivatives are understood to be total derivatives. (In the case of particles, only the total time derivative features.) Because any two Lagrangian densities that are equal up to such a four-divergence give rise to the same Euler-Lagrange equations on the application of Hamilton's principle (a point we will return to in the next section and in the Appendix), it is straightforward to show that Noether transformations that leave the action quasi-invariant in turn leave the explicit form of the Euler-Lagrange equations unchanged. It is in this sense that they represent symmetry transformations.\footnote{More detailed examinations of this point are found in Brading and Brown~\cite{bb*} and particularly~\cite{bb**}.} (It is worth clarifying that there are symmetries in this sense that are non-Noetherian symmetries, and for which there may be no associated conserved charges. We shall see examples in section 4.) 

Noether's first 1918 theorem relates to Lie groups of symmetry transformations of the dependent and/or independent variables that depend on a family of $M$ constant parameters $\varepsilon_{a}$ ($a =1, \ldots ,M$). There are in the literature many good accounts of this theorem, generalised to the case of quasi-invariance, for both particles and fields,\footnote{See, for instance, Hill~\cite{hill}, Doughty~\cite{doughty} pp. 219--222, Barbashov and Nesterenko~\cite{russians} and Jos\'{e} and Saletan~\cite{js*} pp. 565--567. There is a misprint in equation (9.34) in the last treatment; the  $\delta$ (variation) symbol is missing in front of $\Phi^{\alpha}$.} and we shall restrict ourselves to giving the result---without proof---for the special case of 'internal' symmetries. Here, the infinitesimal `global' transformations affect only the fields, and not the space-time coordinates. 

If it is assumed that \textit{all} the fields $\psi_{k}$---or at least the ones involved in the symmetry transformations---are dynamical and satisfy Hamilton's principle, then we expect the explicit variation in the action for each of the infinitesimal parameters $\varepsilon_a$ to consist solely of a surface term, equal to the space-time integral of the four-divergence given in equation~\ref{eq:quasi-invariance}. This result can be expressed as a continuity equation:
\begin{equation}
\frac{d\rho_a}{dt} + \mathbf{\nabla}\!\cdot\mathbf{j}_{a} = 0
\label{eq:continuity}
\end{equation}
where
\begin{equation}
\rho_a = \sum_k \frac{\partial\mathcal{L}}{\partial \dot{\psi}_k}\frac{\partial\left(\delta\psi_k\right)}{\partial\varepsilon_a} - \frac{\partial\Lambda^0}{\partial\varepsilon_a},
\label{eq:rho}
\end{equation}
and
\begin{equation}
\mathbf{j}_{a} = \sum_k \frac{\partial\mathcal{L}}{\partial\left(\mathbf{\nabla}\psi_k\right)}\frac{\partial\left(\delta\psi_k\right)}{\partial\varepsilon_a} - \frac{\partial\mathbf{\Lambda}}{\partial\varepsilon_{a}}.
\label{eq:jay}
\end{equation}
The term $\delta\psi_k$ denotes the infinitesimal variation in $\psi_k$ ($\psi_k\to\psi_k + \delta\psi_k$), and it will be recalled that $\Lambda^0$ and $\mathbf{\Lambda}$ are infinitesimal functions.

Now from equation~\ref{eq:continuity} we have using Gauss's theorem
\begin{equation}
\int_{\Omega}\left[\rho_a(t_2) - \rho_a(t_1)\right]d\mathbf{x} + \int^{t_2}_{t_1}\!\!\int_{\Gamma(\Omega)} \mathbf{j}_a\!\cdot\!d\mathbf{n}  = 0
\label{eq:int-cont}
\end{equation}
where $\Gamma(\Omega)$ is the two-dimensional surface bounded by $\Omega$ and $\mathbf{n}$ is the normal to the surface element on this boundary. Now $\Gamma(\Omega)$ grows as $|\mathbf{x}|^2$, so if we assume that $|\mathbf{j}_a|$ decreases faster than $|\mathbf{x}|^{-2}$, then the second integral in equation~\ref{eq:int-cont} can be made arbitrarily small by making $|\mathbf{x}|$ sufficiently large. For large enough $R$, then, one obtains 
\begin{equation}
\frac{d}{dt}\int_R\rho_ad\mathbf{x} \equiv \frac{dQ_a}{dt} = 0
\label{eq:charge}
\end{equation}
where $Q_a$ is the conserved Noether charge associated with the internal symmetry.

\section{Noether charges for the free Schr\"{o}dinger particle}
The standard Lagrangian density for the free particle in non-relativistic quantum mechanics is given by
\begin{equation}
\mathcal{L}_{free} = - \frac{\hbar^{2}}{2m}\mathbf{\nabla}\psi^{*}\!\cdot\!\mathbf{\nabla}\psi + \frac{i\hbar}{2}\left(\psi^{*}\dot{\psi} - \dot{\psi}^{*}\psi\right)     \label{eq:standard-lag}
\end{equation}
where $\psi(\psi^{*})$ is the wavefunction (complex conjugate) associated with the particle. Applying Hamilton's principle for variations in $\psi^{*}$ leads to the well-known free Schr\"{o}dinger equation via equation~\ref{eq:euler-lagrange}:
\begin{equation}
i\hbar\dot{\psi} = - \frac{\hbar^{2}}{2m}\mathbf{\nabla}^{2}\psi.   \label{eq:S-eqn}
\end{equation}
The corresponding equation for $\psi^*$ is obtained by applying Hamilton's principle with respect to variations in $\psi$.

There are two remarkable features of the derivation of this equation from the standard Lagrangian density. 
\begin{enumerate}
\item The fields $\psi$ and $\psi^{*}$ are treated as independent variables, when they clearly are not.\footnote{It is claimed in Doughty~\cite{doughty} pp. 216 and 351, that the trick works because of the hermiticity of the Lagrangian density.}
\item It is \textit{prima facie} surprising that the Euler-Lagrange equation~\ref{eq:S-eqn} is not second order in both spatial and temporal derivatives, given the form of the Lagrangian density. The implication must be that the latter contains redundant dynamical variables.\footnote{See Cohen-Tannoudji \textit{et al}.\ \cite{cdg} p. 157.}
\end{enumerate}

An arguably more transparent application of Hamilton's principle in this case involves not the usual wavefunction and its complex conjugate but the real fields $\psi_{R}$ and $\psi_{I}$, where
\begin{equation}
\psi = \psi_{R} + i\psi_{I}\:;\hspace{.3in}\psi^* = \psi_{R} -i\psi_{I}.   \label{eq:r&i}
\end{equation}
The second term in the standard Lagrangian density in equation~\ref{eq:standard-lag} can be rewritten in terms of these real fields:
\begin{equation}
\frac{i\hbar}{2}\left(\psi^*\dot{\psi} - \dot{\psi}^*\psi\right) = \hbar\left(\dot{\psi}_R\psi_I - \psi_R\dot{\psi}_I\right).  \label{eq:new-var}
\end{equation}
If we then similarly re-arrange the first term and add the total time derivative
\begin{equation}
\frac{d}{dt}\left(\hbar \psi_R\psi_I\right) = \frac{d}{dt}\left[\frac{\hbar}{4i}\left(\psi^{2} - (\psi^*)^{2}\right)\right]    \label{eq:time-deriv}
\end{equation}
we obtain an equivalent Lagrangian density
\begin{equation}
\mathcal{L}'_{free} = - \frac{\hbar^{2}}{2m}\left[\left(\mathbf{\nabla}\psi_R\right)^{2} + \left(\mathbf{\nabla}\psi_I\right)^{2}\right] + 2\hbar\dot{\psi}_R\psi_I.    \label{eq:new-lag}
\end{equation}

Note that this Lagrangian density depends on the time derivative of $\psi_R$ but not of $\psi_I$.\footnote{See Cohen-Tannoudji \textit{et al}.\ ~\cite{cdg} p. 159.} 
 (We could have subtracted the derivative in equation~\ref{eq:time-deriv} instead of adding it, thereby reversing the role of $\psi_R$ and $\psi_I$.)

It is seen from equation~\ref{eq:euler-lagrange} that the Euler-Lagrange equation defined with respect to variations of $\psi_R$ is
\begin{equation}
\hbar\dot{\psi}_I - \frac{\hbar^{2}}{2m}\mathbf{\nabla}^{2}\psi_R = 0,   \label{eq:S-eqn1}
\end{equation}
and that for variations of $\psi_I$ is
\begin{equation}
\hbar\dot{\psi}_R + \frac{\hbar^{2}}{2m}\mathbf{\nabla}^{2}\psi_I = 0.   \label{eq:S-eqn2}
\end{equation}
Appropriate linear combinations of equation~\ref{eq:S-eqn1} and equation~\ref{eq:S-eqn2} give rise to the usual Schr\"{o}dinger equation~\ref{eq:S-eqn} and its conjugate counterpart.

\subsection{First symmetry}
Let us now consider the global infinitesimal transformations
\begin{equation}
\psi_R \rightarrow \psi'_R = \psi_R + \varepsilon_R \:;\hspace{.3in}\psi_I \rightarrow \psi'_I = \psi_I + \varepsilon_I
\label{eq:transf}
\end{equation}
where the arbitrary parameters $\varepsilon_{R(I)}$ do not depend on space or time. Note that in this case the number $M$ of parameters  equals the number $N$ of fields: both indices $k$ and $a$ range over the set \{$R,I$\}. From equations ~\ref{eq:new-lag} and ~\ref{eq:transf} we have that  $\Lambda^0 = 2\hbar\varepsilon_I\psi_R$ and $\mathbf{\Lambda} = 0$. It can now readily be confirmed from equations~\ref{eq:rho}, ~\ref{eq:jay} and~\ref{eq:new-lag} that the equations of continuity~\ref{eq:continuity} for $a=R$ and $a=I$ have precisely the form of the Euler-Lagrange equations~\ref{eq:S-eqn1} and \ref{eq:S-eqn2} respectively. The Noether equations of continuity in this case coincide with the equations of motion.

Assuming that $\mid\!\! \nabla\psi_{R(I)}\!\! \mid$ decreases faster than $\mid \!\!\mathbf{x} \!\!\mid^{-2}$, we obtain for sufficiently large $\Omega$ our two conserved Noether charges
\begin{equation}
Q_{R(I)} = \int_\Omega\psi_{R(I)}d\mathbf{x}.
\label{eq:charges}
\end{equation}
The existence of these charges implies that 
\begin{equation}
Q = \int_\Omega\psi d\mathbf{x}.
\label{eq:charges'}
\end{equation}
and its complex conjugate are also conserved charges. This result is to be compared with the more familiar derivation of the conserved Noether charge $\int\! \left|\psi\right|^{2}\!d\mathbf{x}$ resulting from the internal symmetry associated with global gauge (phase) tranformations of the wavefunction for the general Schr\"{o}dinger equation (involving possible external potentials).\footnote{For this derivation, see, for instance, Doughty~\cite{doughty} pp. 222--223.} This charge is of course real and gauge independent; the charges~\ref{eq:charges} are real but gauge dependent and charge~\ref{eq:charges'} is complex and gauge dependent.\footnote{Gauge dependent Noether charges are familiar in the case of non-abelian gauge fields in relation to the global SU(2) symmetry; see Barbashov and Nesterenko~\cite{russians}, section 12.}

It should be noted finally that the above results can also be obtained by applying Noether's theorem in the case of the standard Lagrangian \ref{eq:standard-lag} and the transformation
\begin{equation}
\psi \rightarrow \psi' = \psi + \varepsilon
\label{eq:transf'}
\end{equation}
and its conjugate,\footnote{The local transformation $\psi(\mathbf{x},t) \to \psi'(\mathbf{x},t) = \alpha(\mathbf{x},t)\psi(\mathbf{x},t) + \beta(\mathbf{x},t)$ is mentioned in an unworked exercise in Doughty~\cite{doughty}, p. 223, designed to show that the full Schr\"{o}dinger equation---with scalar but no vector potential---has no local gauge freedom in $\psi$. Doughty does not refer to the special case above of $\alpha = 0$ and $\beta$ global in relation to the free particle dynamics, nor the case of $\beta = 0$ and $\alpha$ global, which we discuss briefly in section 4 below.} where $\varepsilon = \varepsilon_{R} + i \varepsilon_{I}$, etc. Here the Noether equations of continuity are just the usual Schr\"{o}dinger equation \ref{eq:S-eqn} and its conjugate counterpart. (In the Appendix, we examine other possible forms of the Lagrangian density, including complex-valued ones, for the free particle.) In this case it is more obvious that the internal transformations do not preserve the square integrability of the wavefunction. But this is no impediment to applying the machinery of Noether's theorem. Indeed the symmetry~\ref{eq:transf'} is a non-relativistic analogue of the  `translation' symmetry for the real massless scalar field studied in connection with the spontaneous breaking of symmetry in relativistic field theory---a Noetherian symmetry which is not unitarily implementable (see section 5 below).

\subsection{Second symmetry}
Consider now the infinitesimal transformations
\begin{equation}
\psi_R \rightarrow \psi'_R = \psi_R + \vec{\varepsilon}_R\cdot\mathbf{x} \:;\hspace{.3in}\psi_I \rightarrow \psi'_I = \psi_I + \vec{\varepsilon}_I\cdot\mathbf{x}
\label{eq:transf''}
\end{equation}
where the arbitrary vectors $\vec{\varepsilon}_{R(I)}$ again do not depend on space or time. In this case, the parameter index $a$ ranges over the six values of the ordered pairs ~\{$R(I), j$\}, and we have that $\Lambda^0 = 2\hbar \psi_R \vec{\varepsilon_I}\cdot\mathbf{x}$ and $\mathbf{\Lambda} = \frac{-\hbar^2}{m}\left(\psi_R \vec{\varepsilon}_R + \psi_I \vec{\varepsilon}_I\right)$. The density and 3-current associated with the choice of parameter $\varepsilon_{Rl}$, for example, are, given equations~\ref{eq:rho} and ~\ref{eq:jay},
\begin{equation}
\rho^{Rl} = 2\hbar \psi_I x_l\:;\hspace{.3in} j^{Rl}_k = \frac{-\hbar^2}{m}\left(\frac{\partial\psi_R}{\partial x_k}x_l -\psi_R\delta^k_l\right).
\label{eq:rho-jay}
\end{equation}
These expressions, when applied in~\ref{eq:continuity}, lead to an equation of continuity which, as the reader will quickly confirm, reduces to the Euler-Lagrange equation~\ref{eq:S-eqn1}. The corresponding argument based on the choice of parameter $\varepsilon_{Il}$ likewise leads to the equation~\ref{eq:S-eqn2}. But now our Noether charges are, for $\Omega$ sufficiently large, 
\begin{equation}
Q^{R(I)}_k = \int_\Omega\psi_{R(I)}x_k d\mathbf{x}\:;\hspace{.2in}Q_k = \int_\Omega\psi x_k d\mathbf{x}\:;\hspace{.2in}Q^*_k = \int_\Omega\psi^*x_k d\mathbf{x}
\label{eq:charges2}
\end{equation}
if $j^{R(I)l}_k$ in ~\ref{eq:rho-jay} falls off sufficiently fast with distance.

We note finally that as with the first symmetry, all these results are also derivable by considering, in relation to the standard Lagrangian~\ref{eq:standard-lag}, the symmetries for $\psi$ and $\psi^*$ that are the consequences of~\ref{eq:transf''}.

\section{Aside: non-Noetherian continuous symmetries}
Consider multiplication of the wavefunction (complex conjugate) by a complex constant $c$ ($c^*$). The form of the general Schr\"{o}dinger equation is clearly unaffected by this transformation, but the standard Lagrangian density~\ref{eq:standard-lag} is not quasi-invariant in the sense of equation ~\ref{eq:quasi-invariance}. Rather, the transformed Lagrangian density is just the original times $|c|^2$; or $\delta\mathcal{L}_{free} = (|c|^2 - 1)\mathcal{L}_{free}$. The transformed Lagrangian density is equivalent to the original in the sense that the Euler-Lagrange equations are insensitive to such multiplicative constants in the Lagrangian. But no continuity equation is implied by the existence of this kind of internal symmetry, unless $|c| = 1$---which is the  familiar case of global gauge transformations.

It might be of interest to note that the symmetry introduced in the last paragraph corresponds to the transformations
\begin{equation}
\psi_R \rightarrow \psi'_R = c_R\psi_R -c_I\psi_I \:;\hspace{.3in}\psi_I \rightarrow \psi'_I = c_R\psi_I + c_I\psi_R,
\label{eq:transf'''}
\end{equation}
where $c = c_R + ic_I$, etc. Under these transformations, $\delta\mathcal{L}'_{free} = (c_R^2 + c_I^2 - 1)\mathcal{L}'_{free}$ up to a divergence, where $\mathcal{L}'_{free}$ is given in equation~\ref{eq:new-lag}.

A final remark in this context concerns the possible generalization of the global transformation~\ref{eq:transf'} to a local transformation, in which $\varepsilon$ depends on the space-time location. It is obvious that this transformation is no longer a symmetry in relation to the Schr\"{o}dinger equation~\ref{eq:S-eqn}. Consider, however, the amusing case where the function $\varepsilon$ is itself a solution of this equation. Then if $\psi$ in ~\ref{eq:transf'} is itself a solution, so is $\psi'$: this is just the superposition principle! Indeed the linearity of the Schr\"{o}dinger equation (free or otherwise) ensures that superposing is a symmetry transformation.  

Does this mean that we can apply to quantum mechanics the lessons of Noether's second 1918 theorem, which is concerned with the connection between local symmetries and certain off-shell `identities'?\footnote{For discussions of Noether's second theorem see, e.g., Trautman~\cite{traut} and Brading and Brown~\cite{bb*} and ~\cite{bb**}.} Not in this case. As in the previous example of an internal symmetry, the superposition symmetry is not a Noetherian symmetry; the Lagrangian is not quasi-invariant under the transformation.

\section{Analogous Noetherian symmetries in electrodynamics}
For systems of relativistic fields $\phi_k$ in Minkowski space-time associated with the Lagrangian density $\mathcal{L}$, the relativistic generalizations of equations~\ref{eq:continuity} and ~\ref{eq:rho}, ~\ref{eq:jay} are
\begin{equation}
\partial_\mu j^{\mu a} = 0\:;\hspace{.3in} j^{\mu a} = \sum_k\frac{\partial\mathcal{L}}{\partial(\partial_{\mu}\phi_k)} \frac{\partial\left(\delta \phi_k\right)}{\partial\varepsilon_a} - \frac{\partial\Lambda^\mu}{\partial\varepsilon_a}
\label{eq:rel-jay}
\end{equation}
(We henceforth use the Einstein convention for summing over repeated Greek indices; $\mu = 0, 1, 2, 3$). In the case of the massless scalar field $\phi$ with ~$\mathcal{L} = \frac{1}{2}\partial_\mu\phi\partial^\mu\phi$, the infinitesimal global transformation
\begin{equation}
\phi (x) \to \phi'(x) = \phi(x) + \varepsilon
\label{eq:transf-rel}
\end{equation}
is a Noetherian symmetry with $\Lambda^\mu = 0$. Such non-unitarily implementable symmetries are called ``hidden'' or ``spontaneously broken''; in this particular case the continuity equation once again coincides with the equation of motion, $\partial_\mu\partial^\mu\phi = 0$.\footnote{See Aitchison~\cite{ait}, sections 5.4, 5.5.} We have seen a non-relativistic analogue for a complex field in section 3.1; there is a relativistic analogue for vector fields in classical electromagnetism, as we now see.

The standard Lagrangian density of the source-free electromagnetic field in Minkowski space-time is
\begin{equation}
\mathcal{L}_{EM} = - \frac{1}{4}F_{\mu\nu}F^{\mu\nu}
\label{eq:em-lag}
\end{equation}
where $F_{\mu\nu} = \partial_{\mu}A_\nu -\partial_\nu A_\mu$ and $A_\mu$ is the electromagnetic 4-potential. As is well-known, applying Hamilton's principle with respect to variations in $A_{\mu}$ leads to the field equations
\begin{equation}
\partial_\mu F^{\mu\nu} = 0
\label{eq:maxwell}
\end{equation}
Let us now consider an infinitesimal transformation in the potential analogous to the transformations~\ref{eq:transf}, ~\ref{eq:transf'} and~\ref{eq:transf-rel}
\begin{equation}
A_\mu \to A'_\mu = A_{\mu} + \varepsilon_\mu
\label{eq:A-transf}
\end{equation}
where $\varepsilon_\mu$ is an arbitrary infinitesimal constant co-vector field. Under this transformation, the Lagrangian~\ref{eq:em-lag} is obviously strictly invariant ($\Lambda_\mu = 0$). 

Now from~\ref{eq:rel-jay} we have
\begin{equation}
\partial_\mu j^{\mu\nu} = 0
\label{eq:rel-continuity}
\end{equation}
and 
\begin{equation}
j^{\mu\nu} = \frac{\partial\mathcal{L}_{EM}}{\partial(\partial_{\mu}A_{\sigma})} \frac{\partial\left(\delta A_{\sigma}\right)}{\partial\varepsilon_{\nu}} = -\frac{1}{4}F^{\mu\sigma}\delta^{\nu}_{\sigma} = -\frac{1}{4}F^{\mu\nu}.
\label{eq:rel-jay}
\end{equation}
So it is seen that again the equations of continuity coincide with the equations of motion~\ref{eq:maxwell}.

For $\nu = 1, 2, 3$, these equations have as their 1+3 formulation
\begin{equation}
\frac{\partial\mathbf{E}}{\partial t} + \nabla\times\mathbf{B} = 0
\label{eq:3-maxwell}
\end{equation}
where $\mathbf{E}$ and $\mathbf{B}$ are the electric and magnetic field vectors respectively. Integrating this equation over space, and recalling
\begin{equation}
\int_\Omega\nabla\times\mathbf{B}d\mathbf{x} = \int_{\Gamma(\Omega)}\mathbf{B}\times d\mathbf{n}
\end{equation}
then it is seen that $\int_\Omega\mathbf{E}d\mathbf{x}$ is conserved over time for $\Omega$ sufficiently large, if $\mathbf{B}\to 0$ as $|\mathbf{x}|\to\infty$. It should be noted that these boundary conditions are highly non-trivial for the propagating free field; in general there is no Noether charge associated with the symmetry~\ref{eq:A-transf}.

For the electromagnetic analogue of the second quantum symmetries~\ref{eq:transf''} consider the following equivalent transformations
\begin{eqnarray}
F_{\mu\nu} \to F'_{\mu\nu} &=& F_{\mu\nu} + \varepsilon_{\mu\nu} \\
A_{\mu} \to A'_{\mu} &=& A_{\mu} + \frac{1}{2}\varepsilon_{\nu\mu}x^{\nu}
\label{eq:em-transf'}
\end{eqnarray}
where $\varepsilon_{\mu\nu}$ is an arbitrary constant antisymmetric second-rank tensor field. In this case, the electromagnetic Lagrangian density~\ref{eq:em-lag} is quasi-invariant, and it can easily be shown that for infinitesimal $\varepsilon_{\mu\nu}$, $\Lambda^{\mu} = \frac{1}{4} A_{\nu}\varepsilon^{\mu\nu}$. In this case the parameter index $a$ corresponds to the brace of indices $\sigma$, $\rho$ and we have as the equation corresponding to~\ref{eq:rel-jay}
\begin{eqnarray}
j^{\mu\sigma\rho} &=& \frac{\partial\mathcal{L}_{EM}}{\partial(\partial_{\mu}A_{\nu})} \frac{\partial\left(\delta A_{\nu}\right)}{\partial\varepsilon_{\sigma\rho}} - \frac{\partial \Lambda^{\mu}}{\partial\varepsilon_{\sigma\rho}} \\
&=& \frac{1}{2}F^{\mu\rho}x^\sigma - \frac{1}{2}F^{\mu\sigma}x^\rho -\eta^{\mu\sigma}A^\rho + \eta^{\mu\rho}A^{\sigma},
\label{eq:rel-jay2}
\end{eqnarray}
where $\eta^{\mu\nu}$ is the Minkowski metric tensor. Given that $F_{\mu\nu}$ is built out of first derivatives of $A_\mu$, the structural similarity with equation~\ref{eq:rho-jay} is striking. We see at any rate that these four-currents are gauge dependent,  and that now the equations of continuity 
\begin{equation}
\partial_\mu j^{\mu\sigma\rho} = \frac{1}{2}\left[\left(\partial_\mu F^{\mu\rho}\right)x^\sigma - \left(\partial_\mu F^{\mu\sigma}\right)x^\rho\right] = 0
\label{eq:continuity''}
\end{equation}
which follow from the quasi-invariance of the Lagrangian density do not coincide with the field equations~\ref{eq:maxwell}.

\section{Appendix: Alternative Lagrangians}
Besides the free Lagrangian densities that we have already encountered in equations~\ref{eq:standard-lag} and \ref{eq:new-lag}, we might consider a further five distinct but equivalent ones. We get these from the standard one~\ref{eq:standard-lag} by allowing the first term to be replaced by either $(\frac{\hbar^{2}}{2m})\psi\mathbf{\nabla}^{2}\psi^{*}$ or $(\frac{\hbar^{2}}{2m})\psi^{*}\mathbf{\nabla}^{2}\psi$, and allowing the second term to be replaced by $-i\hbar\dot{\psi}^{*}\psi$. All these Lagrangian densities differ from one another by scalar terms of the form $\mathbf{\nabla}(\cdot)$ and/or $d/dt(\cdot)$. All give rise to the same Euler-Lagrange equations associated with variations in $\psi$ and $\psi^{*}$. 

Some remarks related to specific cases may be helpful to the reader.

(i) Consider two of these Lagrangian densities
\begin{eqnarray}
\tilde{\mathcal{L}} & = & \frac{\hbar^2}{2m}\psi\mathbf{\nabla}^2\psi^* - i\hbar\dot{\psi}^*\psi \\
\bar{\mathcal{L}} & =& \frac{\hbar^2}{2m}\psi\mathbf{\nabla}^2\psi^*  + \frac{i\hbar}{2}\left(\psi^*\dot{\psi} - \dot{\psi}^*\psi \right)
\label{eq:two-lags}
\end{eqnarray}
Given that $\tilde{\mathcal{L}}$ and $\bar{\mathcal{L}}$ depend on second order space derivatives, the Euler-Lagrange equation defined relative to variations in $\psi$, for example, is now (compare with ~\ref{eq:euler-lagrange})
\begin{equation} 
		\frac{\partial \mathcal{L}}{\partial\psi} - \mathbf{\nabla}\cdot\frac{\partial \mathcal{L}}{\partial (\mathbf{\nabla}\psi)} + \mathbf{\nabla}^2\frac{\partial \mathcal{L}}{\partial (\mathbf{\nabla}^2 \psi)} - \partial_{t}\frac{\partial \mathcal{L}}{\partial \dot{\psi}} = 0.  \label{eq:euler-lagrange'}
\end{equation} 
This equation follows from Hamilton's principle relative to arbitrary variations in $\psi$ only if it assumed that the variations both in $\psi$ and its first space derivatives vanish on the boundary $\Gamma(\Sigma)$.

(ii) It is striking that the Schr\"{o}dinger equation~\ref{eq:S-eqn}, which  is only second order in space derivatives, is derivable from $\tilde{\mathcal{L}}$ and $\bar{\mathcal{L}}$, which are second order in these derivatives. This is reminiscent of the situation in the general theory of relativity concerning the Hilbert Lagrangian density $R\sqrt{g}$, where $R$ is the curvature scalar and $g$ the determinant of the metric field $g^{\mu\nu}$. This functional depends on second order derivatives of $g^{\mu\nu}$, but gives rise to the Einstein field equations that contain no higher than second derivatives. A simple way to see how this can happen is by considering Einstein's own lesser-known 1916 gravitational Lagrangian density (the so-called $\Gamma\!\!-\!\!\Gamma$ Lagrangian), which is first-order. Both actions give rise to the same Euler-Lagrangian equations, because they differ only by a total divergence. In relation to Einstein's action, the second-order terms in Hilbert's action \textit{are all contained in this total divergence}.\footnote{For a fuller discussion of this point, see Brown and Brading~\cite{bb}.} In quantum mechanics the standard free Lagrangian density in equation~\ref{eq:standard-lag} is related to each of $\tilde{\mathcal{L}}$ and $\bar{\mathcal{L}}$ in the same way.

Note that these cases provide counterexamples to a common claim in the literature concerning dynamically equivalent Lagrangians, which has ramifications for the notion of quasi-invariance of the Lagrangian. The claim is that two Lagrangian densities differing by a total divergence are dynamically equivalent only when they are of the same order in the derivatives---so that when a Lagrangian density is first order, and a divergence is added to it, the resulting Lagrangian density is dynamically equivalent only when the argument of the divergence can contains no derivatives of the fields.\footnote{This claim is found in, for example, Hill~\cite{hill}, Doughty~\cite{doughty} pp. 178, 218, Cohen-Tannoudji \textit{et al}.\ ~\cite{cdg} p. 83, and Jos\'{e} and Saletan~\cite{js}, pp. 67--68. (In relation to the last treatment, it is correctly pointed out in the comments on p. 68 that are bounded by vertical rules that it is not necessary for two Lagrangians to yield the same dynamics that they differ by a divergence term; but the preceding argument in the same section seems to contradict this.)} It is also worth noting that each of $\tilde{\mathcal{L}}$ and $\bar{\mathcal{L}}$ and the density to be introduced in the next remark are complex-valued, in defiance of the occasional claim that the action must be real-valued.\footnote{For this claim see Cohen-Tannoudji \textit{et al}.\ ~\cite{cdg} p. 87.}

(iii) The standard Lagrangian density in ~\ref{eq:standard-lag} is only quasi-invariant with respect to the complex conjugate of the global transformation in equation~\ref{eq:transf'}, but it is clearly the case that $\tilde{\mathcal{L}}$ is invariant, and the same goes for another of the above-mentioned densities
\begin{equation}
\hat{\mathcal{L}} = - \frac{\hbar^2}{2m}\mathbf{\nabla}\psi^*\nabla\psi - i\hbar\dot{\psi}^*\psi. 
\label{eq:hat-lag}
\end{equation}
The situation is partly reversed with respect to the Galilean coordinate transformations (boosts), which are of course a symmetry of the Schr\"{o}dinger equation~\ref{eq:S-eqn}.\footnote{For a recent discussion of the Galilean covariance of quantum mechanics see Brown and Holland~\cite{bh}.}  It can be shown that the standard Lagrangian density is strictly Galilean invariant, but $\hat{\mathcal{L}}$ in equation~\ref{eq:hat-lag}, for instance, is only quasi-invariant. It is recalled that the standard Lagrangian for a free non-relativistic classical particle $m\dot{q}^2/2$ is likewise only quasi-invariant with respect to Galilean boosts.\footnote{An unfamiliar example of quasi-invariance under spatial translations for a classical non-relativistic particle in free fall is found in Jos\'{e} and Saletan~\cite{js*}, p. 128.}

\section{Acknowledgements}
We wish to thank Ian Aitchison for pointing out the analogy between the symmetry in 3.1 and the internal `translation' symmetry for the relativistic massless scalar field. One of us (H.R.B.) wishes to gratefully acknowledge extensive and enlightening discussions over several years with Katherine Brading, on matters Noetherian.

\end{document}